\documentclass[apj,iop]{emulateapj}
\shorttitle{Cutoff frequencies for transverse tube waves}
\shortauthors{Routh, Musielak, \& Hammer}
\usepackage{apjfonts}
\usepackage{amsmath}
\begin{document}
\title{Global and local cutoff frequencies for transverse\\
waves propagating along solar magnetic flux tubes}

\author{S.\ Routh$^1$, Z.~E.\ Musielak$^{2,3}$, and R.\ Hammer$^3$}
\affil{$^1$Department of Physics, R.V.\ College of Engineering, 
       Bangalore, India} 
\email{swati.routh@lcd.co.edu}
\affil{$^2$Department of Physics, University of Texas
       at Arlington, Arlington, TX  76019, USA}
\email{zmusielak@uta.edu}       
\affil{$^3$Kiepenheuer-Institut f\"ur Sonnenphysik, 
       Sch\"oneckstr.\ 6, Freiburg, D--79104 Germany}
\email{hammer@kis.uni-freiburg.de}       
%
\slugcomment{to appear in ApJ Vol.\ 763 (2013)} 
\received{2012 July 29}
\accepted{2012 December 4}

\begin{abstract}
It is a well-established result that the propagation of linear 
transverse waves along a thin but isothermal magnetic flux tube 
is affected by the existence of the global cutoff frequency, which 
separates the propagating and non-propagating waves.  In this paper, 
the wave propagation along a thin and non-isothermal flux tube is 
considered and a local cutoff frequency is derived. The effects of 
different temperature profiles on this local cutoff frequency are 
studied by considering different power-law temperature distributions 
as well as the semi-empirical VAL C model of the solar atmosphere. 
The obtained results show that the conditions for wave propagation 
strongly depend on the temperature gradients.  Moreover, the local 
cutoff frequency calculated for the VAL C model gives constraints 
on the range of wave frequencies that are propagating in different 
parts of the solar atmosphere.  These theoretically predicted 
constraints are compared to observational data and are used to 
discuss the role played by transverse tube waves in the atmospheric 
heating and dynamics, and in the excitation of solar atmospheric 
oscillations.
\end{abstract}


\section{Introduction}
   
Magnetic flux tubes existing in the photosphere and lower chromosphere 
of the Sun are considered to be narrow bundles of strong magnetic field 
lines that rapidly expand with height in the solar atmosphere (e.g., 
Solanki 1993). The fundamental modes of linear oscillations of these flux 
tubes are typically identified with longitudinal, transverse and torsional 
tube waves (e.g., Defouw 1976; Roberts \& Webb 1978; Roberts 1979, 1981; 
Spruit 1981, 1982; Priest 1982; Hollweg 1985; Roberts 1991; Roberts \& 
Ulmschneider 1997), with the two latter waves being Alfv\'en-like waves.

Observational evidence for the existence of Alfv\'en-like waves in different 
regions of the solar atmosphere was given by high-resolution observations 
performed by the Solar Optical Telescope (SOT) and the X-Ray Telescope (XRT) 
onboard the Hinode Solar Observatory.  According to De Pontieu et al.\ (2007) 
and Cirtain et al.\ (2007), signatures of Alfv\'en waves were observed by the 
SOT and XRT instruments, respectively.  Moreover, Alfv\'en waves were also 
reported by Tomczyk et al.\ (2007), who used the Coronal Multi-Channel 
Polarimeter of the National Solar Observatory.  Interpretations of these 
observations were given by Van Doorsselaere et al.\ (2008) and Antolin et
al.\ (2009), who concluded that the reported observational results describe
kink waves.  As of today, the most convincing observational evidence for 
the existence of Alfv\'en waves in the quiescent solar atmosphere was 
given by McIntosh et al.\ (2011), who reported indirect evidence for such 
waves found in observations by Solar Dynamic Observatory (SDO).  

Moreover, Fujimura \& Tsuneta (2009) used SOT observations to study 
fluctuations in pores and intergranular magnetic structures, and concluded 
that such fluctuations could be explained by the existence of longitudinal 
(sausage) and transverse (kink) waves propagating along magnetic flux tubes 
embedded in the solar photosphere.  They found oscillation periods of 3-6 min 
for the pores and 4-9 min for the intergranular magnetic elements.  More 
recently, Okamoto and De Pontieu (2011) used data from SOT and presented 
observational evidence for the existence of high-frequency transverse waves 
propagating along spiclues in the solar atmosphere.  Similar high-frequency 
tranverse waves were reported by Yurchyshyn et al.\ (2012), who observed type
II spicules using the New Solar Telescope. 

There is also observational evidence for the existence of torsional Alfv\'en 
waves in the solar atmosphere as reported by Jess et al.\ (2009), who interpreted 
data obtained with high spatial resolution by the Swedish Solar Telescope (SST).  
Alfv\'en-like motions were also observed by Bonet et al.\ (2008), who found vortex 
motions of G band bright points around downflow zones in the solar photosphere, 
and by Wedemeyer-B\"ohm \& Rouppe van der Voort (2009), who used data from the 
SST to demonstrate that more disorganized relative motions of photospheric bright 
points can also induce swirl-like motions in the solar chromosphere.  More recently, 
observational evidence for Alfv\'en-like waves was found by Bonet et al.\ (2010), 
who used the ballon-borne Sunrise Telescope, and by De Pontieu et al.\ (2012), who 
used the SST to find torsional motions in spicules.  Moreover, magnetic swirls in 
the solar atmosphere were reported by Wedemeyer-B\"ohm et al.\ (2012). 

Among the above described observational results, the observations perfomed by 
Fujimura \& Tsuneta (2009), Okamoto and De Pontieu (2011), and Yurchyshyn et al.\ 
(2012) are the most relevant because the authors directly refer to transverse tube 
waves, which are the main topic of this paper.  Our goal here is to determine the 
propagation conditions for these waves by calculating the so-called cutoff frequencies.  
In general, there are global cutoff frequencies, which are the same along the entire 
length of the tube, and local cutoff frequencies that are height dependent.  If, 
under certain simplifying approximations, the cutoff frequency is constant along 
the entire length of the tube, it is called a global cutoff.  Otherwise, if it is 
height dependent, it is a local cutoff.  In Section 6, we compare these cutoff 
frequencies to the observational results.   

The global cutoff frequency represents the natural frequency of linear 
oscillations of the magnetic flux tubes, and its value restricts the wave 
propagation to only those frequencies that are higher than this cutoff.  
For isothermal and thin flux tubes, the global cutoff frequencies for 
longitudinal and transverse tube waves were first determined by Defouw 
(1976) and Spruit (1981), respectively.  These global cutoff frequencies 
are ratios of the characteristic wave speeds to the pressure (density) 
scale height. Since the scale height and the wave speeds are constant 
along isothermal and exponentially expanding magnetic flux tubes, the 
resulting cutoffs are the same along the entire length of the tubes.  
The fact that there is no global cutoff frequency for torsional waves 
propagating along isothermal and thin magnetic flux tubes was demonstrated 
by Musielak, Routh \& Hammer (2007). 

A method to derive the global cutoff frequency for longitudinal tube waves 
was introduced by Rae \& Roberts (1982) and Musielak et al.\ (1987, 1995), 
who demonstrated that the wave equation for these waves can be transformed 
into its standard form (also referred to as the Klein-Gordon equation), 
which directly displays the global cutoff frequency.  A similar method was 
used by Musielak \& Ulmschneider (2001, 2003) to determine the global cutoff 
frequency for transverse tube waves.  These global cutoff frequencies are 
important in studies of atmospheric oscillations (e.g., Roberts 1991; Hasan 
\& Kalkofen 1999; Hasan 2003; Musielak \& Ulmschneider 2003; Hasan 2008) 
and chromospheric and coronal heating (e.g., Narain \& Ulmschneider 1996; 
Ulmschneider \& Musielak 2003).
 
The existence of global cutoff frequencies is restricted to 
thin and isothermal magnetic flux tubes. However, in more general 
cases when the flux tubes are either wide and isothermal, or thin 
and non-isothermal, or wide and non-isothermal, the resulting 
cutoff frequencies depend on atmospheric height, which means that 
they are local quantities (e.g., Routh, Musielak \& Hammer 2007). 
A method to determine such local cutoff frequencies was developed 
by Musielak, Musielak \& Mobashi (2006).  This method requires that 
wave equations are cast in their standard forms (i.e.,without terms with 
the first-order derivatives) and that the oscillation theorem (e.g., 
Kahn 1990) is used to obtain the cutoff frequencies. 

The form of the oscillation theorem used in this paper is that given 
by Kahn (1990).  Different forms of this theorem were considered by 
mathematicians depending on specific differential equations and the 
applied boundary conditions.  The basic idea was originally introduced 
by Sturm (1836), who considered a second-order ordinary differential 
equation written in its standard form and established a comparison 
criterion, which allowed him to determine whether solutions to the 
equation were oscillatory or not, without formally solving the 
equation.  Another comparison criterion was developed by Kneser 
(1893) and later generalized by Fite (1918), Hille (1948), Wintner 
(1949, 1957), and many others.  The most important generalization 
of Kneser's criterion was done by Leighton (1950, 1962).  In more 
recent work, some oscillation theorems were established for linear 
(e.g., Li \& Yeh 1995, 1996) and nonlinear (e.g., Li 1998; Lee et 
al.\ 2005; Tyagi 2009) second-order differential equations.

The oscillation theorem presented by Kahn (1990), which is used in 
this paper, is equivalent to Kneser's criterion, which is also known as 
Kneser's theorem (e.g., Bohner \& \"Unal 2005).  First applications of 
this theorem to solar physics problems were done by Musielak \& Moore 
(1995), who considered the propagation of linear Alfv\'en waves in an 
isothermal solar atmosphere.  Then, Schmitz \& Fleck (1998) used the 
theorem to establish criteria for acoustic wave propagation in the 
solar atmosphere.  More recently the theorem was used by Schmitz \& 
Fleck (2003), Musielak et al.\ (2006, 2007), Routh et al.\ (2007, 2010), 
and Hammer et al.\ (2010).                   

In this paper, we use this method to derive the cutoff frequency 
for transverse waves propagating along a thin and non-isothermal 
magnetic flux tube embedded in the solar atmosphere. The effects 
of temperature gradients on the cutoff frequency are studied for 
several power-law temperature models as well as for the reference 
mean solar atmosphere model C given by Vernazza, Avrett \& Loeser 
(1981). The height dependence of the cutoff frequency in these 
models is calculated, and it is shown that the value of this cutoff 
at a given atmospheric height determines the frequency that transverse 
tube waves must exceed in order to be propagating at this height. 
The results are compared to those previously obtained for a thin 
and isothermal magnetic flux tube. We also briefly discuss 
implications of our results for the energy and momentum balance 
of the solar atmosphere.

Our paper is organized as follows: the governing equations and 
derivation of our wave equations are given in Section 2; the global 
cutoff frequency for linear transverse waves propagating along a 
thin and isothermal magnetic flux tube is obtained in Section 3; the 
local cutoff frequency for the wave propagation along a thin and 
non-isothermal magnetic flux tube is derived in Section 4; the local 
cutoff frequencies for different power-law temperature distributions 
and for the VAL C solar atmosphere model are presented and discussed 
in Sections 5 and 6, respectively; and conclusions are given in Section 7.  

\section{Wave equations}

We consider a thin and non-isothermal magnetic flux tube that is 
embedded in the solar atmosphere.  The tube axis is assumed to have 
a circular cross-section and an unperturbed tube axis oriented 
vertically along the $z$-axis, so that gravity $\vec g = - g \hat z$. 
The tube density, pressure, temperature, and magnetic field are 
respectively given by $\rho_0 = \rho_0 (z)$, $p_0 = p_0 (z)$, $T_0 
= T_0(z)$, and $B_0 = B_0 (z)$.  Moreover, the density, pressure and 
temperature of the external non-magnetic ($B_e = 0$) atmosphere are 
represented by $\rho_e = \rho_e (z)$, $p_e = p_e (z)$ and $T_e = T_e(z)$, 
respectively.  We assume that the tube is in thermal equilibrium with 
its surroundings, which means that at each atmospheric height $T_0 (z) 
= T_e (z)$.

In the thin flux tube approximation (e.g., Ferriz-Mas et al.\ 1989),
the horizontal pressure balance for the tube becomes
\begin{equation}
p_0 + {{B_{0}^2} \over {8 \pi}} = p_e\ .
\label{s2eq1}
\end{equation} 

The equation for transverse linear oscillations of the tube was originally 
derived by Spruit (1981), who obtained
\begin{equation}
{{\partial^2 \xi} \over {\partial t^2 }} - c_\mathrm{k}^{2}(z)\ {{\partial^2 \xi} 
\over {\partial z^2 }} - g\frac{\rho_{0} - \rho_e}{\rho_{0} + \rho_e} 
{{\partial \xi} \over {\partial z }} \  = 0\ ,
\label{s2eq2}
\end{equation}

\noindent
where $\xi$ is the horizontal displacement of the tube, and $c_\mathrm{k}$ is the 
characteristic wave speed given by 
\begin{equation}
c_\mathrm{k} (z) = {{B_0} \over {\sqrt{4 \pi (\rho_0 + \rho_e)}}}\ .
\label{s2eq3}
\end{equation}

The relationship between the displacement $\xi$ and the corresponding magnetic 
field perturbation $b_x$ was obtained by Stix (1991) and Bogdan et al.\ (1996), 
and it can be written as  
\begin{equation}
b_x = B_{0}\frac{\partial \xi}{\partial z}\ . 
\label{s2eq4}
\end{equation}

Defining the velocity perturbation 
\begin{equation}
v_x =\frac{\partial \xi}{\partial t}\ , 
\label{s2eq5}
\end{equation}

\noindent
and using 
\begin{equation}
g\frac{\rho_{0} - \rho_e}{\rho_{0} + \rho_e}=-\frac{c_\mathrm{k}^{2}}{2H}\ , 
\label{s2eq6}
\end{equation}

\noindent
where $H = c_\mathrm{S}^2 / \gamma g$ is the pressure scale height, with $\gamma$ 
being the ratio of specific heats and $c_\mathrm{S}$ being the speed of sound given 
by $c_\mathrm{S} = \sqrt{\gamma p_0/\rho_0}$, we write Equation (\ref{s2eq2}) in the 
following form
\begin{equation}
{{\partial^2 v_{x}} \over {\partial t^2 }} - c_\mathrm{k}^{2}(z)\ {{\partial^2 v_{x}} 
\over {\partial z^2 }} + \frac{c_\mathrm{k}^{2}(z)}{2H(z)} {{\partial v_{x}} 
\over {\partial z }} \  = 0\ .
\label{s2eq7}
\end{equation}

Then we combine Equations (\ref{s2eq2}) and (\ref{s2eq4}), use $dB_0/dz = -B_0/2H$, 
and obtain
\begin{equation}
{{\partial^2 b_x} \over {\partial t^2 }} - c_\mathrm{k}^2(z)\ {{\partial^2 b_x} 
\over {\partial z^2 }} - c_\mathrm{k}^2(z) \left [ {2 \over c_\mathrm{k}(z)} \left ( 
{{dc_\mathrm{k}(z)} \over {dz}} \right ) + {1 \over {2H(z)}} \right ] 
{{\partial b_x} \over {\partial z}} = 0\ .
\label{s2eq8}
\end{equation}

\noindent
The derived wave equations describe the propagation of linear transverse waves 
along a thin and non-isothermal magnetic flux tube.  Since the wave equations 
for $v_x$ and $b_x$ are different, there is a phase difference between the wave 
variables.  Physical consequences of the existence of this phase shift will be 
explored in the next sections.   

\section{Global cutoff frequency}

We now consider the special case of a thin and isothermal magnetic flux tube by taking 
$T_0$ = const, $T_e$ = const and $T_0 = T_e$, which gives $c_\mathrm{k}$ = const and $H$ = 
const. This allows us to write the wave equations (see Equations (\ref{s2eq7}) and 
(\ref{s2eq8})) as
\begin{equation}
{{\partial^2 v_x} \over {\partial t^2 }} - c_\mathrm{k}^2\ 
{{\partial^2 v_x} \over {\partial z^2 }} + {{c_\mathrm{k}^2} 
\over {2 H}}\ {{\partial v_x} \over {\partial z }}\ = 0
\label{s3eq1}
\end{equation}
and 
\begin{equation}
{{\partial^2 b_x} \over {\partial t^2 }} - c_\mathrm{k}^2\ 
{{\partial^2 b_x} \over {\partial z^2 }} - {{c_\mathrm{k}^2} 
\over {2 H}}\ {{\partial b_x} \over {\partial z }}\ = 0\ .
\label{s3eq2}
\end{equation}

\noindent
Comparison of the above wave equations shows that they are 
different, which means that the behavior of the wave variables 
$v_x$ and $b_x$ is different even for a thin and isothermal 
flux tube. 

The wave equation for $v_x$ was originally obtained by Spruit
(1981, 1982) and he showed that the propagation of transverse 
waves along a thin and isothermal magnetic flux tube is affected 
by a cutoff frequency.  A new result presented in this paper is 
that we also derive the wave equation for $b_x$, show that it is 
different than the one obtained for $v_x$, and demonstrate how to 
obtain the cutoff frequency for both wave variables. 

To obtain the cutoff frequency, we follow Musielak \& Ulmschneider 
(2001) and cast Equations (\ref{s3eq1}) and (\ref{s3eq2}) into their 
standard (or Klein-Gordon) forms (e.g., Roberts 1981; Rae \& Roberts 
1982).  The required transformations are $v_x (z,t) = v_1 (z,t) 
\rho_0^{-1/4}(z)$ and $b_x (z,t) = b_1(z,t) \rho_0^{1/4}(z)$, and 
the resulting wave equations can be written as
     \begin{equation}
\left [ {{\partial^2} \over {\partial t^2 }} - c_\mathrm{k}^2 
{{\partial^2} \over {\partial z^2 }} + \Omega_\mathrm{k}^2 
\right ]\ [v_1 (z,t), b_1 (z,t)]\ = 0\ ,
     \label{s3eq3}
     \end{equation}
where
     \begin{equation}
\Omega_\mathrm{k} = {{c_\mathrm{k}} \over {4 H}}
     \label{s3eq4}
     \end{equation}

\noindent
is called here Spruit's cutoff frequency. Since $c_\mathrm{k}$ = const 
and $H$ = const 
in the isothermal case, 
$\Omega_\mathrm{k}$ is also constant, which means that 
it is a global quantity. 

With the coefficients of Equation (\ref{s3eq3}) being constant, we 
can make Fourier transforms in time and space, and derive the 
global dispersion relation: $(\omega^2 - \Omega_\mathrm{k}^2) = k^2 c_\mathrm{k}^2$, 
where $\omega$ is the wave frequency and $k = k_z$ is the wave 
vector along the tube axis. Based on this dispersion relation, 
$\Omega_\mathrm{k}$ is the global cutoff frequency for transverse tube 
waves. According to the dispersion relation, the waves are 
propagating when $\omega > \Omega_\mathrm{k}$ and $k$ is real, and 
non-propagating when either $\omega = \Omega_\mathrm{k}$ with 
$k = 0$ or $\omega < \Omega_\mathrm{k}$ with $k$ being imaginary; in 
the latter case, the waves are called evanescent waves. 

According to Musielak et al.\ (2006, 2007) and Petukhov (2006),
gradients of restoring forces, which can introduce characteristic 
scale heights, are responsible for the origin of cutoff frequencies 
for different waves.  Using this argument, Musielak et al.\ (2006) 
argued about the origin of the acoustic cutoff frequency and Musielak
et al.\ (2007) showed that the propagation of torsional tube waves is 
cutoff-free.  Moreover, Petukhov (2006) demonstrated that horizontal, 
non-uniform but potential magnetic fields have no influence on a cutoff 
frequency for magneto-acoustic waves and, as a result, the cutoff is 
identical to Lamb's cutoff frequency, $\Omega_\mathrm{S} = c_\mathrm{S} / 2H$, which was 
obtained by Lamb (1908, 1975) for acoustic waves propagating in a 
stratified and isothermal medium.  

The results obtained here for transverse tube waves clearly show that 
the vertical and non-uniform (exponentially diverging) magnetic field 
of a thin and isothermal flux tube is responsible for the origin of 
Spruit's cutoff frequency.  Actually, the existence of this cutoff 
is caused by the horizontal pressure balance, which relates the tube 
magnetic pressure to the gas pressure.  Since the latter introduces 
the pressure scale height $H$ that also determines the rate with which 
the field diverges with the atmospheric height, the tube magnetic field 
and its characteristic scale height are the main physical reasons for the 
existence of Spruit's cutoff frequency.

Finally, we may relate the Spruit cutoff $\Omega_\mathrm{k}$ to the Lamb cutoff
$\Omega_\mathrm{S}$ by writing 
\begin{equation}
\Omega_\mathrm{k} = {1 \over 2} {{c_\mathrm{k}} \over {c_\mathrm{S}}} \Omega_\mathrm{S}\ ,     
\label{s3eq5}
\end{equation}

\noindent
which shows that the ratio of $c_\mathrm{k}$ to $c_\mathrm{S}$ determines how much $\Omega_\mathrm{k}$
differs from $\Omega_\mathrm{S}$.  For the special case $c_\mathrm{k} = c_\mathrm{S}$, we have 
$\Omega_\mathrm{k} = \Omega_\mathrm{S} / 2$.  Hence, $\Omega_\mathrm{k}$ can be much larger than 
$\Omega_\mathrm{S}$ when $c_\mathrm{k} >> c_\mathrm{S}$ and much smaller when $c_\mathrm{k} << c_\mathrm{S}$.

\section{Local cutoff frequency}

We now consider a thin but non-isothermal magnetic flux tube
whose internal temperature distribution is $T_0 = T_0 (z)$, 
and assume that at each atmospheric height the tube is in 
thermal equilibrium with its surroundings, which means that 
$T_0 (z) = T_e (z)$. The presence of temperature gradients 
makes both $c_\mathrm{k}$ and $H$ to be functions of $z$, and this 
leads to wave equations for $v_x$ and $b_x$ that have 
non-constant coefficients (see Equations (\ref{s3eq1}) and (\ref{s3eq2})).

Since these wave equations cannot be solved by using Fourier 
transforms in space, a different method to obtain cutoff 
frequencies is required.  Such a method was originally developed 
by Musielak et al.\ (2006) for acoustic waves propagating in 
non-isothermal media.  Routh et al.\ (2007, 2010) and Hammer et 
al.\ (2010) used the method to determine cutoff frequencies for 
the propagation of torsional waves in thick (isothermal) and 
thin (non-isothermal) magnetic flux tubes.  Here, this method 
will be used to derive local cutoff frequencies for 
transverse waves propagating along a thin flux tube with 
different temperature profiles. 

The method is based on the oscillation theorem given by Kahn 
(1990), which follows the original work of Sturm (1836) and 
Kneser (1893).  Actually, there are numerous oscillation 
theorems developed by mathematicians for various differential 
equations and their boundary conditions (e.g., Swanson 1968;
Teschl 2011, and references therein).  However, most of these
theorems cannot be directly applied to Euler's equation (e.g.,
Wong 1996; Aghajani \& Roomi 2012), which is used in this paper.
Fortunately, solutions to Euler's equations are well-known (e.g.,
Murphy 1960), and they form the basis for our method used in 
this paper.   

\subsection{Transformed wave equations}

The method begins with introducing a new variable     
     \begin{equation}
d \tau = {{dz} \over {c_\mathrm{k} (z)}}\ .
     \label{s5eq1}
     \end{equation}

\noindent
The physical meaning of this variable becomes obvious after 
both sides of the above equation are integrated (see below). 
Then, $\tau (z)$ is the actual wave travel time between a 
height at which a wave source is located and a given height 
$z$ along the axis of the magnetic flux tube. 

We express Equations (\ref{s2eq7}) and (\ref{s2eq8}) in terms of 
the variable $\tau$ and obtain the following transformed 
wave equations 
     \begin{equation}
\left [ {{\partial^2} \over {\partial t^2 }} - {{\partial^2} 
\over {\partial \tau^2 }} + \left ( {{c_\mathrm{k}^{\prime}} \over 
{c_\mathrm{k}}} + {{c_\mathrm{k}} \over {2 H}} \right )  {{\partial} \over 
{\partial \tau }}\right] v_x (\tau, t)\ = 0
     \label{s5eq2}
     \end{equation}
and
     \begin{equation}
\left [ {{\partial^2} \over {\partial t^2 }} - {{\partial^2} 
\over {\partial \tau^2 }} -\left ( {{c_\mathrm{k}^{\prime}} \over 
{c_\mathrm{k}}} + {{c_\mathrm{k}} \over {2 H}} \right ) {{\partial} \over 
{\partial \tau }}  \right ] b_{x} 
(\tau, t)\ = 0\ ,
     \label{s5eq3}
     \end{equation}

\noindent
where $c_\mathrm{k} = c_\mathrm{k} (\tau)$, $c_\mathrm{k}^{\prime} = d c_\mathrm{k} / d \tau$ 
and $H = H(\tau)$. 

\subsection{Standard wave equations}

To convert the transformed wave equations into their standard 
forms, we use
      \begin{equation}
v_x (\tau, t) = v(\tau, t) \exp \left [ + {1 \over 2} 
\int^{\tau} \left ( {{c_\mathrm{k}^{\prime}} \over {c_\mathrm{k}}} + 
{{c_\mathrm{k}} \over {2 H}} \right ) 
d \tilde \tau \right ]
      \label{s5eq4}
      \end{equation}
and
     \begin{equation}
b_{x} (\tau, t) = b(\tau, t) \exp \left [ - {1 \over 2} 
\int^{\tau} \left ( {{c_\mathrm{k}^{\prime}} \over {c_\mathrm{k}}} + 
{{c_\mathrm{k}} \over {2 H}} \right ) d \tilde \tau \right]\ ,
      \label{s5eq5}
      \end{equation}

\noindent
and obtain 
     \begin{equation}
\left [ {{\partial^{2}} \over {\partial t^2}} - {{\partial^2}
\over {\partial \tau^2}} + \Omega_{\mathrm{cr},v}^2 (\tau) \right ] 
v (\tau, t) = 0
     \label{s5eq6}
     \end{equation}
and 
     \begin{equation}
\left [ {{\partial^{2}} \over {\partial t^2}} - {{\partial^2}
\over {\partial \tau^2}} + \Omega_{\mathrm{cr},b}^2 (\tau) \right ] 
b (\tau, t) = 0\ ,
     \label{s5eq7}
     \end{equation}
where
    \begin{equation} 
\Omega_{\mathrm{cr},v}^2 (\tau) = \Omega_\mathrm{k}^2 (\tau)+ {3 \over 4} 
\left ( {{c_\mathrm{k}^{\prime}} \over {c_\mathrm{k}}} \right )^2 - {1 \over 2}
{{c_\mathrm{k}^{\prime \prime}}\over {c_\mathrm{k}}} +\frac{c_\mathrm{k}}{4H}\ 
{{H^{\prime}} \over {H}}
    \label{s5eq8}
    \end{equation}
and
    \begin{equation} 
\Omega_{\mathrm{cr},b}^2 (\tau) = \Omega_\mathrm{k}^2 (\tau) - {1 \over 4} \left( 
{{c_\mathrm{k}^{\prime}} \over {c_\mathrm{k}}} \right )^2 + {1 \over 2}
{{c_\mathrm{k}^{\prime \prime}}\over {c_\mathrm{k}}} - \frac{c_\mathrm{k}}{4H}\ 
{{H^{\prime}} \over {H}} + \frac{c_\mathrm{k}^{\prime}}{2H}\ , 
     \label{s5eq9}
    \end{equation}
\noindent
with $c_\mathrm{k}^{\prime \prime} = d^2 c_\mathrm{k} / d \tau^2$ and $H^{\prime} 
= d H / d \tau$.  The frequencies $\Omega_{\mathrm{cr},v}$ and $\Omega_{\mathrm{cr},b}$ 
are known as the critical frequencies (Musielak et al.\ 1992, 2006).  

\subsection{Turning-point frequencies}

We make the Fourier transform in time $[ v (\tau, t), b (\tau, t)] = 
[\tilde v (\tau), \tilde b (\tau)] e^{- i \omega t}$, where $\omega$ 
is the wave frequency, and write Equations (\ref{s5eq6}) and (\ref{s5eq7}) 
as
     \begin{equation}
\left [ {{d^2} \over {d \tau^2}} + \omega^2 - \Omega_{\mathrm{cr},v}^2 (\tau) 
\right ] \tilde v (\tau) = 0
     \label{s5eq10}
     \end{equation}
and
     \begin{equation}
\left [ {{d^2} \over {d \tau^2}} + \omega^2 - \Omega_{\mathrm{cr},b}^2 (\tau) 
\right ] \tilde b (\tau) = 0\ .
     \label{s5eq11}
     \end{equation}

Using the oscillation theorem given by Kahn (1990) and comparing the 
above equations to Euler's equation (e.g., Murphy 1960), we obtain 
the following conditions for the wave propagation 
     \begin{equation}
\omega^2 - \Omega_{\mathrm{cr},v}^2 (\tau) > {1 \over {4 \tau^2}}
     \label{s5eq12}
     \end{equation}
and 
     \begin{equation}
\omega^2 - \Omega_{\mathrm{cr},b}^2 (\tau) > {1 \over {4 \tau^2}}\ . 
     \label{s5eq13}
     \end{equation}

\noindent
Note that applications of the oscillation theorem to other wave
propagation problems relevant to solar physics were previously 
considered by Musielak \& Moore (1995) and Schmitz \& Fleck (1998); 
see also Routh et al.\ (2010) for more recent results. The idea of 
using Euler's equation to determine turning-point frequencies was 
first introduced by Musielak \& Moore (1995). 

We follow Musielak et al.\ (2006) and define the turning-point 
frequencies as
     \begin{equation}
\Omega_{\mathrm{tp},v}^2 (\tau) = \Omega_{\mathrm{cr},v}^2 (\tau) + {1 \over 
{4 \tau^2}}
     \label{s5eq14}
     \end{equation}
and
     \begin{equation}
\Omega_{\mathrm{tp},b}^2 (\tau) = \Omega_{\mathrm{cr},b}^2 (\tau) + {1 \over {4 
\tau^2}}\ ,
     \label{s5eq15}
     \end{equation}

\noindent
where $\tau$ is the actual wave propagation time (see Routh et al.
2010) and is given by 
     \begin{equation}
\tau (z) = \int^z {{d \tilde z} \over {c_\mathrm{k} (\tilde z)}}\
+ \tau_C\ ,
     \label{s5eq16}
     \end{equation}
\noindent
with $\tau_C$ being an integration constant to be evaluated when 
flux tube models are specified (see Sections 6 and 7). 

It is important to point out that the turning-point frequencies 
were obtained by making the derived wave equations (see Equations (\ref{s5eq10}) 
and (\ref{s5eq11})) equivalent to Euler's equation, so that the 
well-known solutions of the latter can be directly applied to 
these wave equations. 

\subsection{Converting $\tau$ into $z$}

Having obtained the turning-point frequencies (see Equations (\ref{s5eq14}) 
and (\ref{s5eq15})), we now express them in terms of the $z$ variable, 
which requires using Equation (\ref{s5eq16}) to convert $\tau$ to $z$. 

We begin with the critical frequencies $\Omega_{\mathrm{cr},v} (\tau)$ and  
$\Omega_{\mathrm{cr},b} (\tau)$ given by Equations (\ref{s5eq8}) and (\ref{s5eq9}).
Using the following expressions 
     \begin{equation}
{1 \over {c_\mathrm{k}}} {{d c_\mathrm{k}} \over {d \tau}} = {{d c_\mathrm{k}} 
\over {d z}}\ ,
     \label{s5eq17}
     \end{equation}
     \begin{equation}
{1 \over {c_\mathrm{k}}} {{d^2 c_\mathrm{k}} \over {d \tau^2}} = c_\mathrm{k} 
{{d^2 c_\mathrm{k}} \over {d z^2}} + \left ( {{d c_\mathrm{k}} \over 
{d z}} \right )^2\ ,
     \label{s5eq18}
     \end{equation}
and 
     \begin{equation}
{1 \over {c_\mathrm{k}}} {{d H} \over {d \tau}} = {{d H} 
\over {d z}}\ ,
     \label{s5eq19}
     \end{equation}
we obtain

\begin{equation} 
\Omega_{\mathrm{cr},v}^2 (z) = \Omega_\mathrm{k}^2 (z) \left [ 1 + 4 \left ( 
\frac {dH}{dz} \right ) \right ] - {1 \over 2}c_\mathrm{k} \frac {d^{2} 
c_\mathrm{k}} {d z^{2}} + {1 \over 4}\left(\frac{d c_\mathrm{k}}{dz}\right)^2\ ,
    \label{s5eq20}
    \end{equation}

and 
\begin{equation} 
    \label{s5eq21}
\begin{split}
\Omega_{\mathrm{cr},b}^2 (z) = & \, \Omega_\mathrm{k}^2 (z) \left [ 1 - 4 \left ( 
\frac {dH}{dz} \right ) \right ] + \frac{c_\mathrm{k}}{2 H} \frac{dc_\mathrm{k}}
{dz} \\
& + {1 \over 2}c_\mathrm{k}\frac{d^{2} c_\mathrm{k}}{d z^{2}} + {1 \over 4}
\left(\frac{d c_\mathrm{k}}{dz}\right)^2\ ,
\end{split}
\end{equation}

We now use Equation (\ref{s5eq16}) to express the conditions 
for wave propagation and the turning-point frequencies 
as functions of $z$, and obtain 
     \begin{equation}
\left [ \omega^2 - \Omega_{\mathrm{cr},v}^2 (z) \right ] > {1 \over 4} 
\left [ \int^z {{d \tilde z} \over {c_\mathrm{k} (\tilde z)}}\ +
\tau_C \right ]^{-2}
     \label{s5eq22}
     \end{equation}
and 
     \begin{equation}
\left [ \omega^2 - \Omega_{\mathrm{cr},b}^2 (z) \right ] > {1 \over 4} 
\left [ \int^z {{d \tilde z} \over {c_\mathrm{k} (\tilde z)}}\ 
+ \tau_C \right ]^{-2}\ . 
     \label{s5eq23}
     \end{equation}

The turning-point frequencies are
     \begin{equation}
\Omega_{\mathrm{tp},v}^2 (z) = \Omega_{\mathrm{cr},v}^2 (z) + {1 \over 4} 
\left [ \int^z {{d \tilde z} \over {c_\mathrm{k} (\tilde z)}}\ 
+ \tau_C \right ]^{-2}
     \label{s5eq24}
     \end{equation}
and
     \begin{equation}
\Omega_{\mathrm{tp},b}^2 (z) = \Omega_{\mathrm{cr},b}^2 (z) + {1 \over 4} 
\left [ \int^z {{d \tilde z} \over {c_\mathrm{k} (\tilde z)}}\ 
+ \tau_C \right ]^{-2}\ .
     \label{s5eq25}
     \end{equation}

We shall use the turning-point frequencies given above to 
determine the cutoff frequency for transverse tube waves.

\subsection{Cutoff frequency}

We follow Musielak et al.\ (2006) and take the larger of the 
two turning-point frequencies to be the cutoff frequency 
$\Omega_\mathrm{cut}$,
     \begin{equation}
\Omega_\mathrm{cut} (z) = {\max} [\Omega_{\mathrm{tp},v}(z), \Omega_{\mathrm{tp},b} (z)]\ .
     \label{s5eq26}
     \end{equation}

Our selection of $\Omega_\mathrm{cut}$ is physically justified by the 
fact that in order to have propagating transverse tube waves, 
the wave frequency $\omega$ must always be higher than any 
turning-point frequency. In other words, the choice guarantees 
that propagating wave solutions are obtained for both wave 
variables, and that the cutoff frequency does separate the 
propagating and non-propagating wave solutions (see also Hammer 
et al.\ 2010). Hence, the condition for propagating waves is 
$\omega > \Omega_\mathrm{cut}$. Based on our definition of the 
turning-point frequencies, the condition for non-propagating 
(evanescent) waves is $\omega \leq \Omega_\mathrm{cut}$. 

The above results show that the cutoff frequency can only be
determined when a flux tube model is specified, so that the   
turning-point frequencies can be obtained (see Sections 5 and 6).
In addition, one must keep in mind that the conditions given 
by Equation (\ref{s5eq26}) must be checked at each height because 
in some regions along the tube $\Omega_{\mathrm{tp},v}$ could be larger 
than $\Omega_{\mathrm{tp},b}$, and the opposite could be true in other 
regions (see Section 6).    

\section{Models with power-law temperature distributions}

Let us consider the following temperature distribution inside 
the tube 
    \begin{equation} 
T_0 (z) = T_{00} \xi^m\ , 
     \label{s6eq1}
    \end{equation}

\noindent
where $\xi = z / z_0$ is the distance ratio, with $z_0$ being a 
fixed height in the model, $T_{00}$ is the temperature at $z_0$, 
and $m$ can be any real number. Note that in all power-law models 
discussed below
a wave source is 
assumed to be 
located at $\xi = 1$, which means that in all 
calculations $\xi \geq 1$. In addition, for all models 
we take
$z_0 = 10$
km, $T_{00} = 5000$ K, $c_\mathrm{k0} = 10$ km/s, and for gravity we take 
its solar value. The resulting temperature distributions for $m$ 
being a positive integer that ranges from $1$ to $5$ are presented 
in Figure~1.
\begin{figure}
\figurenum{1}
\plotone{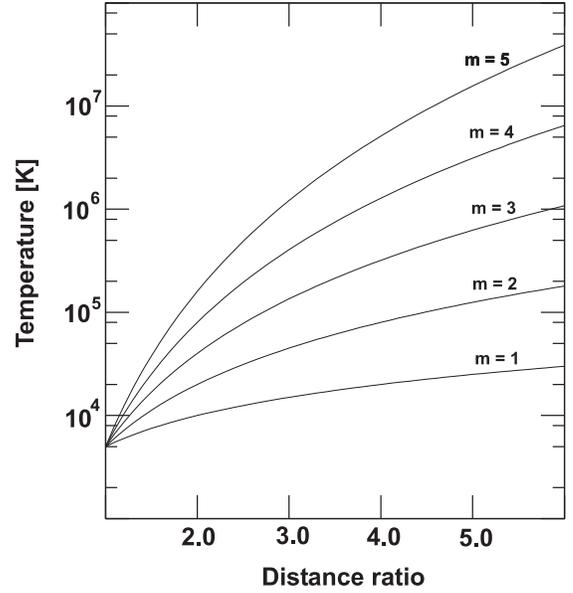}
\caption{Temperature is plotted versus the distance ratio
$z / z_0$ for the power-law temperature models with $m$ 
ranging from $1$ to $5$.}
\end{figure}

\subsection{Case of m = 1}

To describe the process of deriving a local cutoff frequency,
we begin with the simplest case of $m = 1$, which corresponds 
to the temperature varying linearly with $\xi$. We calculate 
$\rho_0$, $p_0$, $B_0$ and $c_\mathrm{k}$ as functions of $\xi$, and 
use Equation (\ref{s5eq16}) to obtain    
    \begin{equation} 
\tau (\xi) = 2 {z_0 \over c_\mathrm{k0}} \xi^{1/2} + \tau_C\ , 
     \label{s6eq2}
    \end{equation}

\noindent
where $c_\mathrm{k0}$ is the value of $c_\mathrm{k}$ at $z_0$ and $\tau_C$ is 
the integration constant. To determine this constant, we assume 
that $\tau (\xi = 1) = \tau_0 \equiv z_0 / c_\mathrm{k0}$, which gives
$\tau_C = - \tau_0$. 
 
We express the turning-point frequencies given by Equations (\ref{s5eq24}) 
and (\ref{s5eq25}) in terms of $\xi$. This gives 
    \begin{equation} 
\Omega_{\mathrm{tp},v}^2 (\xi) = \Omega_\mathrm{k0}^2 \left [ 1 + 4 {{H_{00}} \over 
{z_0}} + \left ( {H_{00} \over z_0} \right )^2 \left ( 3 + g_1 (\xi)
\right ) \right ] \xi^{-1}\ ,
     \label{s6eq3}
    \end{equation}

\noindent
where $\Omega_\mathrm{k0} = c_\mathrm{k0} / 4 H_{00}$, with $H_{00}$ being the scale 
height at $z_0$, and
    \begin{equation} 
g_1 (\xi) = {4\xi \over {(2 \xi^{1/2}- 1)^2}}
     \label{s6eq4}
    \end{equation}
and
    \begin{equation} 
\Omega_{\mathrm{tp},b}^2 (\xi) = \Omega_\mathrm{k0}^2 \left [ 1 - \left ( {{H_{00}} 
\over {z_0}} \right )^2 \left ( 1 - g_1 (\xi) \right ) \right ] \xi^{-1}\ .
     \label{s6eq5}
    \end{equation}

The difference between the turning-point frequencies is 
    \begin{equation} 
{{\Omega_{\mathrm{tp},v}^2 (\xi)- \Omega_{\mathrm{tp},b}^2 (\xi)} \over {\Omega_\mathrm{k0}^2}} 
= 4 \left ( {{H_{00}} \over {z_0}} \right ) \left ( \frac{H_{00}} {z_{0}} 
+1 \right ) \xi^{-1}\ , 
     \label{s6eq6}
    \end{equation}

\noindent
which shows that $\Omega_{\mathrm{tp},v}^2 (\xi)$ is larger than $\Omega_{\mathrm{tp},b}^2 (\xi)$.
Hence, $\Omega_{\mathrm{tp},v}$ becomes the local cutoff frequency, so we write $\Omega_\mathrm{cut} 
(\xi) = \Omega_{\mathrm{tp},v} (\xi)$ and  
    \begin{equation} 
\Omega_\mathrm{cut} (\xi) = \Omega_\mathrm{k0} \left [ 1 + 4 {{H_{00}} \over {z_0}} + \left ( 
{H_{00} \over z_0} \right )^2 \left ( 3 + g_1 (\xi) \right ) \right ]^{1/2} 
\xi^{-1/2}\ .
     \label{s6eq7}
    \end{equation}

\noindent
The local cutoff frequency $\Omega_\mathrm{cut}$ is plotted as a function of $\xi$ 
in Figure 2. It is seen that the cutoff frequency decreases with the atmospheric 
height in the model with $m = 1$.
\begin{figure}
\figurenum{2}
\plotone{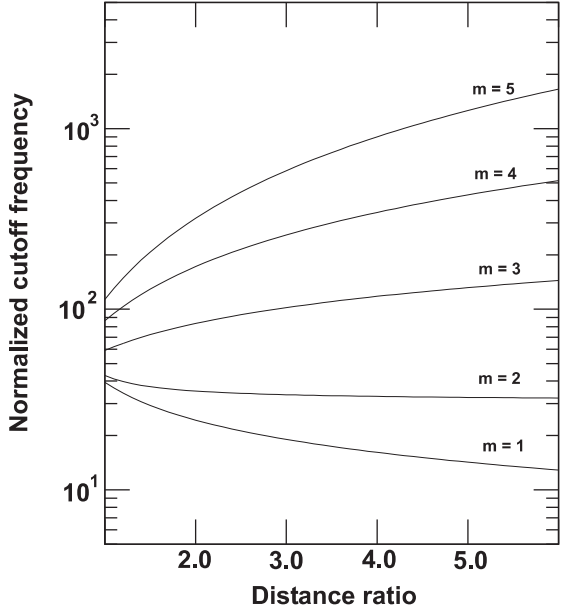}
\caption{The normalized cutoff frequency $\Omega_\mathrm{cut} / 
\Omega_\mathrm{k0}$ is plotted versus the distance ratio $z / z_0$ 
for the power-law temperature models with $m$ ranging 
from $1$ to $5$.}
\end{figure}

\subsection{Case of m = 2}

 In this case, $\tau (\xi)$ is given by 
    \begin{equation} 
\tau (\xi) = {z_0 \over c_\mathrm{k0}} ln \xi + \tau_C\ , 
     \label{s6eq8}
    \end{equation}

\noindent
where $\tau_C$ is the integration constant determined from the 
assumption that $\tau (\xi = 1) = \tau_0 \equiv z_0 / c_\mathrm{k0}$. 
This gives $\tau_C = \tau_0$. 
The turning-point frequencies are   
    \begin{equation} 
\Omega_{\mathrm{tp},v}^2 (\xi) = \Omega_\mathrm{k0}^2 \left [ \xi^{-2} + 8 {{H_{00}} 
\over {z_0}} \xi^{-1} + 4 \left ( {{H_{00}} \over z_0} \right )^2 
\left ( 1 + g_2 (\xi) \right ) \right ] 
     \label{s6eq9}
    \end{equation}
and
     \begin{equation} 
\Omega_{\mathrm{tp},b}^2 (\xi) = \Omega_\mathrm{k0}^2 \left [ \xi^{-2} + 4 
\left ({{H_{00}} \over z_0} \right )^2 \left ( 1 + g_2 (\xi) \right ) 
\right ]\ ,
     \label{s6eq10}
    \end{equation}
where
    \begin{equation} 
g_2 (\xi) = {1 \over {(1 + ln \xi)^2}}\ . 
     \label{s6eq11}
    \end{equation}

\noindent
Here $\Omega_{\mathrm{tp},v}(\xi)$ is always larger than $\Omega_{\mathrm{tp},b}(\xi)$. 
Thus, we choose $\Omega_{\mathrm{tp},v}$ as the local cutoff frequency and write 
$\Omega_\mathrm{cut} (\xi) = \Omega_{\mathrm{tp},v} (\xi)$, or
    \begin{equation} 
\Omega_\mathrm{cut} (\xi) = \Omega_\mathrm{k0} \left [ \xi^{-2} + 8 {{H_{00}} 
\over {z_0}} \xi^{-1} + 4 \left ( {{H_{00}} 
\over z_0} \right )^2 \left ( 1 + g_2 (\xi) \right ) \right ]^{1/2}\ . 
     \label{s6eq12}
    \end{equation}

\noindent
This cutoff frequency is plotted as a function of $\xi$ in Figure 2, which 
shows that the cutoff remains almost constant in the model with $m = 2$.

\subsection{Cases with $m > 2$}

In this general case of $m > 2$, we obtain 
    \begin{equation} 
\tau (\xi) = {z_0 \over c_\mathrm{k0}} \left ( {2 \over {2 - m}} \right ) 
\xi^{1 - m/2} + \tau_C\ , 
     \label{s6eq13}
    \end{equation}

\noindent
where the integration constant $\tau_C$ is evaluated by taking $\tau 
(\xi = 1) = \tau_0 \equiv z_0 / c_\mathrm{k0}$; note that our choice of $\tau_0$ 
gives the same physical parameters at $z = z_0$ for all the power-law models. 
After evaluating $\tau_C$, we calculate

\begin{equation} 
\label{s6eq14}
\begin{split}
\Omega_{\mathrm{tp},v}^2 (\xi) = \Omega_\mathrm{k0}^2 
        \left[\vphantom{\left(\frac{1_2}{1_2}\right)^2} \xi^{-m} \right.
   &\, + 4 m \left(\frac{H_{00}}{z_0} \right) \xi^{-1} \\
   &\, + m (4-m) \left( \frac{H_{00}}{z_0} \right)^2 \xi^{m-2} \\
   &\, + 4 \left( \frac{m-2}{m} \right)^2 \left( \frac{H_{00}}{z_0}
          \right)^2 g_3 (\xi) 
        \left.\vphantom{\left(\frac{1_2}{1_2}\right)^2}\right] \ ,
\end{split}
\end{equation}
where
    \begin{equation} 
g_3 (\xi) = \left ( 1 + {2 \over m} \xi^{1-m/2} \right )^{-2}\ ,
     \label{s6eq15}
    \end{equation}
and
\begin{equation} 
\label{s6eq16}
\begin{split}
\Omega_{\mathrm{tp},b}^2(\xi) = \Omega_\mathrm{k0}^2
     \left[\vphantom{\left(\frac{1_2}{1_2}\right)^2} \xi^{-m} \right. 
            & + m(3m-4)\left(\frac{H_{00}}{z_0}\right)^2 \xi^{m-2} \\
            & + 4\left(\frac{m-2}m \right)^2 
                     \left(\frac{H_{00}}{z_0}\right)^2  g_3(\xi) 
     \left.\vphantom{\left(\frac{1_2}{1_2}\right)^2}\right] \ .
\end{split}
\end{equation}

We now calculate the difference between these turning-point frequencies 
and obtain
\begin{equation}
{{\Omega_{\mathrm{tp},b}^2 (\xi)- \Omega_{\mathrm{tp},v}^2 (\xi)} \over {\Omega_\mathrm{k0}^2}} 
= 4 m (m-2)\left ( {{H_{00}} \over {z_0}} \right )^2 \xi^{m-2} - 4 
m \frac{H_{00}}{z_0} \xi^{-1}\ .
     \label{s6eq17}
    \end{equation}

\noindent
Since in this model $ H_{00} > z_{0}$, $m > 2$, and $\xi > 1$, $\Omega_{\mathrm{tp},b}^2$ 
is larger than $\Omega_{\mathrm{tp},v}^2$.   This indicates that the local cutoff frequency 
is $\Omega_\mathrm{cut} (\xi) = \Omega_{\mathrm{tp},b} (\xi)$, or  
\begin{equation} 
\label{s6eq19}
\begin{split}
\Omega_\mathrm{cut}(\xi) = \Omega_\mathrm{k0} 
     \left[\vphantom{\left(\frac{1_2}{1_2}\right)^2} \xi^{-m} \right.
           & + m(3m-4)\left(\frac{H_{00}}{z_0}\right)^2 \xi^{m-2} \\
           & + 4 \left(\frac{m-2}{m}\right)^2 
                    \left(\frac{H_{00}}{z_0}\right)^2 g_3(\xi)
     \left. \vphantom{\left(\frac{1_2}{1_2}\right)^2}\right]^{1/2}\ .
\end{split}
\end{equation}

\noindent
The cutoff frequency calculated for the power-law temperature models with $m = 3$, 
$4$ and $5$ is plotted versus the distance ratio in Figure 2. It is seen that this 
cutoff frequency always increases with the atmospheric height in the models with 
$m > 2$ and that its increase is much faster for higher values of $m$.  

\subsection{Discussion}

The effects of different temperature gradients on the cutoff frequency for transverse 
tube waves are presented in Figure 2. Since the cutoff frequency is a local quantity, 
its value at a given atmospheric height determines the frequency that the waves must 
exceed in order to be propagating waves at this height. Our results demonstrate that the 
conditions for wave propagation strongly depend on the temperature profiles. 

If the temperature increases linearly with height ($m = 1$), the cutoff frequency 
 starts with its maximum at $z = z_0$ and then decreases with height. For the temperature 
model with $m = 2$, the cutoff frequency remains practically constant with height.
However, the cutoff frequency always increases with height in the temperature models 
with $m \geq 3$; the higher the value of $m$, the steeper an increase of the local 
cutoff frequency with height is observed. 

The main purpose of using the power-law temperature models was to demonstrate 
the dependence of the local cutoff frequency on the increasing steepness of the 
temperature models. Obviously, the power-law models do not properly describe the 
temperature gradients in the solar atmosphere. Therefore, we now consider a more 
realistic model of the solar atmosphere.     

\section{VAL model of the solar atmosphere}

We assume that a thin flux tube is embedded in the reference mean
solar atmosphere model VAL-C (Vernazza et al.\ 1981).  In this
model, the height $z = 0$ km corresponds to unity optical depth,
where the temperature is $6420$ K.  Beyond the temperature minimum
$T_{min} = 4170$ K, which is located at $z = 515$ km, the model
extends through the chromosphere well into the transition region to
the corona, up to a temperature of $4.47 \times 10^5$ K. At the base
of the transition region the VAL-C model exhibits a plateau with a
temperature around $2 \times 10^4$ K, caused by Ly$\,\alpha$ emission.
This plateau was later shown to disappear if the physics of the transition
region is treated more realistically (Fontenla et al.\ 1990). Therefore we
restrict ourselves to the height range from $z = z_0 = 0$ km (optical depth
unity) to the just below the plateau ($z = 2\,113$ km).
 
To calculate the characteristic wave speed $c_\mathrm{k}$ as a function of $z$,
we have to know the magnetic field $B_0(z)$, which is not given in the
VAL-C model. Since the tube is 
assumed to be in 
in thermal equilibrium with its
surroundings at each atmospheric height, we use the local value of the
pressure scale height, which is the same outside and inside the tube,
to evaluate $B_0 (z)$; the calculation begins with 
an assumed surface magnetic field 
$B_0(z_0) = 1500$ G (see Equation~(\ref{s2eq1})). Having obtained the
distribution of the tube magnetic field with height, we then calculate
the gas pressure and density inside the tube, and the characteristic
wave speed $c_\mathrm{k}(z)$ (see Equation~(\ref{s2eq3})).  In 
the lower panel of 
Figure~3 we plot $c_\mathrm{k}$ 
versus the atmospheric height in the model; for comparison we
also plot the sound speed $c_\mathrm{S}$.  The results show that
$c_\mathrm{k}$ is smaller than $c_\mathrm{S}$ in almost the entire model, except in the
upper chromosphere and lower transition region, where $c_\mathrm{k}$ becomes
comparable to, or even slightly larger than, $c_\mathrm{S}$.
\begin{figure*}    
\figurenum{3}
\epsscale{.62}     
\plotone{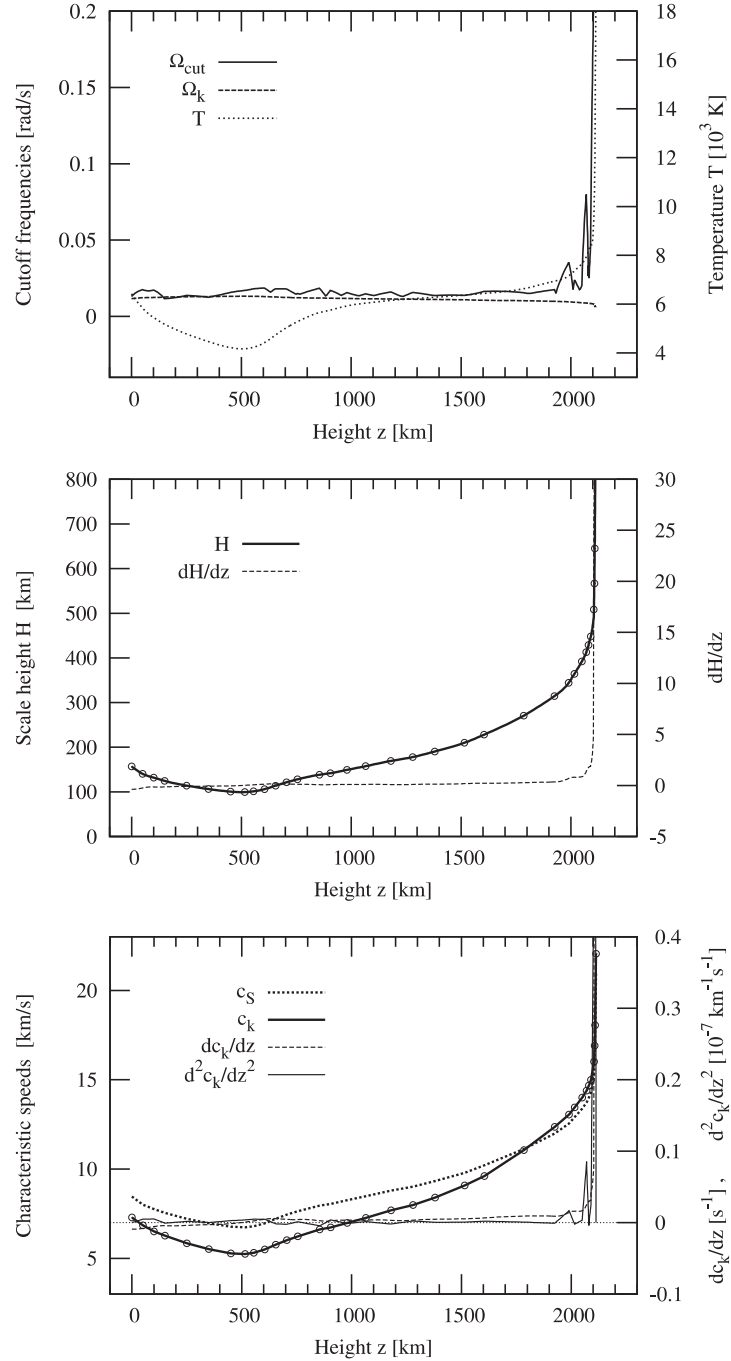}
\caption{The \textit{upper panel} shows the height dependence of the
  cutoff frequency $\Omega_\mathrm{cut}$ for the VAL-C model, for which the
  temperature stratification is shown as a dotted line scaled with the
  right y axis.  For comparison, Spruit's isothermal cutoff frequency
  $\Omega_\mathrm{k} = c_\mathrm{k} / 4 H$ is plotted by using local values of the tube
  speed $c_\mathrm{k}$ and the scale height $H$.  The \textit{middle} and
  \textit{lower} panels, respectively, show $H$ and $c_\mathrm{k}$ as well as
  the sound speed $c_\mathrm{S}$, with scales on the left y axes. The values of
  $H$ and $c_\mathrm{k}$ at the tabulated points of the VAL-C model are marked
  with circles.  The derivatives of the spline curves through these
  points are shown with scales on the right y axes.}
\end{figure*}

In order to calculate the wave propagation time $\tau$, we use
Equation~(\ref{s5eq16}).  For the integration constant $\tau_C = \tau (z =
0) = \tau_0$ we choose a value of 100 s, thus assuming that the waves
have traveled for 100 s before they reach the base $z_0$ of the model.
For typical wave speeds (cf.\ lower panel of Figure~3) this means that
the waves are generated about two scale heights below $z_0$.

To determine the cutoff frequency $\Omega_\mathrm{cut}(z)$ according to
Section 4.5, 
we calculate both turning-point frequencies 
$\Omega_{\mathrm{tp},v} (z)$ and $\Omega_{\mathrm{tp},b}(z)$ and 
select the larger one as $\Omega_\mathrm{cut} (z)$ (see Equations 
(\ref{s5eq20}), (\ref{s5eq21}), and (\ref{s5eq24})--(\ref{s5eq26})).  To 
perform these calculations, we must evaluate the first derivative 
of $c_\mathrm{k}$ and $H$, and the second derivative of $c_\mathrm{k}$.  These 
calculations have to be done numerically because the model consists 
of tabulated data.  
The upper panel of Figure~3 shows the resulting cutoff frequency 
$\Omega_\mathrm{cut} (z)$ as well as Spruit's cutoff frequency $\Omega_\mathrm{k} 
(z) = c_\mathrm{k}(z) / 4 H(z)$ (cf.\ Equation (\ref{s3eq4})), where the latter 
is again treated here as a local, height-dependent quantity.

A striking property of Figure~3 is that the cutoff frequency for the
nonisothermal case, $\Omega_\mathrm{cut} (z)$, is wiggly.  This is mostly
caused by the dependence of the turning-point frequencies on the
second derivative of the tube speed $c_\mathrm{k}$ (cf.\ Equations (\ref{s5eq20})
--(\ref{s5eq21})). Since the VAL-C model is specified by a number of 
tabulated distinct points, the curves in Figure~3 must be plotted by 
finding interpolating or fitting functions through these points and 
determining the first derivative of the scale height, and the first 
and second derivatives of the tube speed, in order to calculate the terms 
in Equations (\ref{s5eq20})--(\ref{s5eq21}).  Such derivatives of interpolated 
curves tend to show a wiggly behavior.  As shown in the middle and
lower panels of Figure~3, these wiggles are stronger near the
temperature minimum and in particular at the foot of the transition
region, where the gradients change rapidly (see also Fawzy \& Musielak
2012).

One could avoid the wiggles by 
approximating the data by simple functions with known smooth derivatives.  
There are also automatic algorithms that can smooth out such wiggles more 
or less completely (e.g., Chartrand 2011).  However, we show them here 
explicitly in order to illustrate the dependence of the cutoff frequency, 
via these derivatives, on the nature of the solar atmosphere, which in 
reality is less homogeneous than the VAL-C model and in addition temporally
variable.  Therefore, we chose the well-proven algorithm by Reinsch (1967), 
which fits the data points by a natural cubic spline, thus giving consistent 
first and second derivatives.  A moderate amount of smoothing can be applied, 
so that the spline needs not go exactly through the points.

Since the value of $\Omega_\mathrm{cut}$ is different at each atmospheric height, 
the frequency $\omega$ of a transverse tube wave must be higher than the 
cutoff at a given height $z$ in order to be propagating at this height.  
Overall, the new cutoff frequency $\Omega_\mathrm{cut}$ is seen to be slightly 
higher than $\Omega_\mathrm{k}$ throughout the chromosphere.  This is mostly 
caused by the fact that the chromospheric temperature increases with 
height, leading to a positive first derivative of the tube speed $c_\mathrm{k}$, 
which increases both turning-point frequencies (cf.\ Equations (\ref{s5eq20}) 
and (\ref{s5eq21})) and thus also their maximum, the cutoff frequency 
$\Omega_\mathrm{cut}$ (Equation (\ref{s5eq26})).

Another difference between the Spruit cutoff and our new result lies
in the behavior at the foot of the transition region.  Here $\Omega_\mathrm{k}$
decreases 
with height 
(which becomes more evident if one continues the calculation
to higher levels of the VAL-C model), while $\Omega_\mathrm{cut}$ increases
rapidly as we approach the steep temperature rise in the transition
region.  Formally this means that a wave with arbitrary frequency
tends to lie above $\Omega_\mathrm{k}$ at the transition region (implying
propagation in the case of an isothermal treatment of the cutoff), but
below our value (implying reflection off the temperature gradient in
our nonisothermal treatment).
This may have implications for the contribution of such transverse
waves to the heating and dynamics of the overlying corona.  However,
we put not much weight on such differences in the high chromosphere
and transition region, fow two reasons. First, we have not provided a
complete study of the reflection properties of the transition region,
which would require a more complex analysis of the complete solar atmosphere.  And
second, a basic approximation underlying all these theories breaks
down much earlier, already in the mid chromosphere - namely, that the
flux tubes can be approximated as thin.  The thin flux tube
approximation, which is the basis for the results presented in this
paper and for Spruit's results as well, is no longer valid at those
heights.  Thus the results obtained for the upper atmosphere must be
taken with caution.  We have extended the calculations up to the foot
of the transition region in order to determine the approximate
behavior of the cutoffs in these layers, even though this is beyond
the formal limit of validity of our analytical results, because
the presented results demonstrate that the temperature gradient in the
upper solar chromosphere and in the solar transition region will have
a major influence on the propagation of transverse tube waves.

The increase of temperature with height requires atmospheric heating,
which is typically identified with acoustic-gravity and flux tube
waves, including transverse tube waves, or with phenomena related to
magnetic reconnection (e.g., Priest 1982; Narain \& Ulmschneider 1996;
Ulmschneider \& Musielak 2003). Our results presented in Figure~3 give
constraints on the range of frequencies of transverse tube waves that
are propagating in different parts of the solar atmosphere.  

Our results show that in the upper photosphere and lower chromosphere 
$\Omega_\mathrm{cut}^\mathrm{phot} \approx 0.01 \,\mathrm{rad/s}$, or the cutoff period is 500 
s, and that in the middle and upper chromosphere $\Omega_\mathrm{cut}^\mathrm{chrom} 
\approx 0.02 \,\mathrm{rad/s}$, which corresponds to the cutoff period of 350 s.  
As already mentioned in Section 1, Fujimura \& Tsuneta (2009) observed 
transverse tube waves with periods of 240-540 s, or the corresponding 
frequency range $\omega \approx (0.01 - 0.03) \,\mathrm{rad/s}$, in intergranular 
magnetic elements.  Comparison of these observational results to our 
theoretical cutoff frequencies shows that most of the observed waves 
are freely propagating in the solar photosphere and lower chromosphere
because $\omega > \Omega_\mathrm{cut}^\mathrm{phot}$, however, in the middle and upper 
parts of the solar chromosphere the waves with frequencies $\omega < 
\Omega_\mathrm{cut}^\mathrm{chrom}$ are not propagating.

Now, the high frequency waves with periods of 45 s, or $\omega \approx 0.1 
\,\mathrm{rad/s}$, observed by Okamoto \& De Pontieu (2011) along spicules do satisfy 
the condition $\omega > \Omega_\mathrm{cut}^\mathrm{chrom}$ and they are obviously freely 
propagating waves in the solar chromosphere.  Similarly, the high frequency 
oscillations with periods ranging from 30 s to 180 s, which corresponds to 
the frequency range $\omega \approx (0.03 - 0.2) \,\mathrm{rad/s}$, detected by 
Yurchyshyn et al.\ (2012) in type II spicules also satisfy the condition 
$\omega > \Omega_\mathrm{cut}^\mathrm{chrom}$, and therefore they are freely propagating 
waves in the solar chromosphere.    

The above constraints can be used to assess the role of transverse 
tube waves in the heating and dynamics of the solar atmosphere.  As 
discussed by Hasan \& Kalkofen (1999), Musielak \& Ulmschneider (2003), 
and Hasan (2003), transverse tube waves may be responsible for the 
excitation of solar atmospheric oscillations observed in magnetically 
active regions near sunspots. The results obtained in this paper can 
be used to determine the natural frequency of the solar atmosphere 
inside thin and non-isothermal magnetic flux tubes and the effects 
of temperature gradients on solar atmospheric oscillations.   

\section{Conclusions}

Our method to determine the cutoff frequency for transverse waves
propagating along a thin and non-isothermal flux tube requires that 
integral transformations are used to cast the wave equations for 
both wave variables in the standard forms. Then the conditions for 
the existence of propagating wave solutions are established by 
using the oscillation theorem. The theorem is also used to obtain 
the turning-point frequency for each wave variable. The larger 
among the two turning-point frequencies is selected as the cutoff
frequency.

To study the effects of temperature gradients on the cutoff frequency, 
we used different power-law temperature distributions.  For the temperature 
that increases linearly with height, the cutoff frequency is large at the 
bottom of the model and then decreases with height.  For the temperature 
profile with the power of $2$, the cutoff frequency remains practically 
constant with height.  However, for powers higher than $2$, the cutoff 
frequency always increases with height; and the higher the power, the 
steeper the increase of the cutoff frequency with height.

We also calculated the cutoff frequency $\Omega_\mathrm{cut}$ as a function
of height $z$ in the VAL-C model and compared it to Spruit's global
cutoff frequency $\Omega_{S}$ that was treated as a height-dependent
quantity. The comparison shows that $\Omega_\mathrm{cut}$ exceeds $\Omega_{S}$, 
and that the differences are especially prominent in the upper parts of 
the model, where however the thin flux tube approximation is not valid 
any longer. The differences in the lower atmosphere, where the thin flux 
tube approximation is valid, may be important for the energy carried by 
transverse tube waves from the solar convection zone, where the waves are 
generated, to the overlying solar atmosphere, where the wave energy is 
deposited.

The cutoff frequency $\Omega_\mathrm{cut}$ calculated for the VAL-C model gives 
constraints on the range of frequencies of transverse tube waves that are 
propagating in different parts of the solar atmosphere.  We compared 
$\Omega_\mathrm{cut}$ to the range of frequencies determined from observations by 
Fujimura \& Tsuneta (2009), Okamoto \& De Pontieu (2011), and Yurchyshyn et 
al.\ (2012).  Based on this comparison, we established that most of the waves 
observed by Fujimura \& Tsuneta (2009) in intergranular magnetic elements are 
freely propagating in the solar photosphere and lower chromosphere, however, 
in the middle and upper solar chromosphere the lower frequency part of these 
waves corresponds to non-propagating waves.  On the other hand, all the waves 
detected by Okamoto \& De Pontieu (2011) and Yurchyshyn et al.\ (2012) in 
spicules are freely propagating waves in the solar chromosphere.    

These constraints are important for understanding the role played by
transverse tube waves in the heating of the solar atmosphere and in
the acceleration of plasma, e.g.\ in spicules and the solar wind. Our
results can be used to determine the natural frequency of the solar
atmosphere inside thin and non-isothermal flux tubes and to study the
effects of the temperature stratification on the excitation of
atmospheric oscillations inside solar magnetic flux tubes.

\acknowledgments

Z.E.M.\ acknowledges the support of this work by the Alexander von 
Humboldt Foundation.

\vspace{1cm}
%

\end{document}